\documentclass[11pt]{article}
\usepackage{geometry}
\usepackage{url,hyperref}
\usepackage{orcidlink}
\usepackage{cite}
\usepackage{amsmath,amssymb,amsfonts}%
\usepackage{amsthm}%
\usepackage{mathrsfs}%
\usepackage[title]{appendix}%
\usepackage{xcolor}%
\usepackage{textcomp}%
\usepackage{manyfoot}%
\usepackage[T1]{fontenc}

\newcommand{\fant}[1]{\phantom{#1}}
\newcommand{\be}{\begin{equation}}
\newcommand{\ee}{\end{equation}}
\newcommand{\wdg}{\wedge}
\newcommand{\ot}{\otimes}
\newcommand{\epsl}{\varepsilon}

\title{The Exterior Calculus of Quadratic Gravity}

\author{Metin Ar\i k\;\orcidlink{https://orcid.org/0000-0001-9512-8581} \footnote{arikm@boun.edu.tr}\\Department of Physics, Faculty of Arts and Sciences,\\ Bo\u gazi\c{c}i University, 34342 Bebek, Be\c{s}ikta\c{s}/\.Istanbul, Turkey
\and
Ahmet Baykal\;\orcidlink{https://orcid.org/0000-0003-0362-9544}\,\footnote{abaykal@ohu.edu.tr}
\\Department of Physics, Faculty of Science, \\Ni\u gde \"Omer Halisdemir  University, 51240 Ni\u gde, Turkey,
\and
Tekin Dereli\;\orcidlink{https://orcid.org/0000-0002-6244-6054} \footnote{tdereli@ku.edu.tr}\\Department of Basic Sciences, Faculty of Engineering and Natural Sciences,\\ Maltepe  University, B\"uy\"ukbakkalk\"oy, 34858 Maltepe/\.Istanbul, Turkey,\\
and\\
Department of Physics, College of Sciences, Ko\c{c} University, \\34450 Sar\i yer/\.Istanbul, Turkey
\and
Taner Tanrıverdi\;\orcidlink{https://orcid.org/0000-0001-8600-4595} \footnote{ttanriverdi@ohu.edu.tr}\\Department of Physics, Faculty of Science, \\Ni\u gde \"Omer Halisdemir  University, 51240  Ni\u gde, Turkey
}

\begin{document}

\maketitle

\begin{abstract}The metric tensor field equations for the general quadratic curvature gravity in four spacetime dimensions are derived by making use of the algebra of the exterior forms defined on pseudo-Riemannian manifolds and the  identities satisfied by the Riemann curvature tensor. The linearized metric field equations are formulated in terms of  perturbation 1-form fields.
\end{abstract}

\section{Introduction}\label{intro}

The classical Einsteinian gravity, assumed to be a low energy limit of a  quantum theory, is to be supplemented by higher powers of curvature terms. In this context, the string theory favors  particular quadratic curvature terms \cite{Boulware-Deser1985} in addition to the scalar curvature. In general, quadratic curvature extension of Einstein gravity has eight degrees of freedom corresponding to a  massive spin-2 ghost and a massive scalar degree of freedom in addition to the healthy massless spin-2  degrees of freedom in the linearized approximation around flat spacetime \cite{stelle1,hindawi,Hinterbichler:2015soa}. The addition of such terms leads to a renormalizable higher curvature gravitational theory
\cite{stelle2,Salvio:2018crh,buchbinder-odintsov-book}. More recently, L\"u and Pope \cite{Lu:2011zk} showed that in an anti-de Sitter vacuum, the massive spin-2 component can be made  massless with a fine-tuned  cosmological constant.

Remarkably, \c{S}entürk et al \cite{Senturk:2012yi,Amsel:2012se,Tekin:2016vli} showed that the perturbative particle spectrum of a gravitational theory governed by the generic Lagrangian  density $F(Riemann)$ can be obtained by constructing a quadratic curvature Lagrangian that has the identical vacuum solutions and the linearized field equations. The linearization of the metric field equations for a gravitational theory as well as its quadratic curvature equivalent is also important in the definition of a total energy for all higher curvature gravitational models.

 A generic quadratic curvature gravity has maximally symmetric vacuum solutions in addition to the flat solution.
A definition of  total energy based on a flux integral was given by Deser and Tekin \cite{deser-tekin1, deser-tekin2,Adami:2017phg} around various  maximally symmetric and flat vacua of the quadratic curvature gravity. Recently, Peng et al \cite{Peng:2023mvd} formulated the conserved quantities in quadratic curvature models with the help of a particular fourth rank tensor and a suitable potential 2-form.

Interestingly, there is a Birkhoff-type theorem \cite{riegert} on the static, spherically symmetric solutions to the conformally invariant quadratic curvature model and  the solution to such a gravitational model is unique.
The pure quadratic curvature models are also interesting in their own right and some geometrical properties and the exact solutions to some particular models are studied in Refs. \cite{kehagias, podolsky}. Recently, in a more efficient approach in studying the algebraically special solutions \cite{Gullu:2011sj,Gurses:2012db,Gurses:2014soa,Malek:2011pn,Pravda:2016fue,Baykal:2014exa,Lu:2015cqa,hjschmidt}, Svarc et al \cite{Svarc:2022fmg} have written down the quadratic curvature gravity equations in terms of the Newman-Penrose scalars.

Previously, Pechlaner and Sexl \cite{sexl,Baykal:2012rr} showed that  a pure quadratic curvature model based on $R^2*1$ does not admit static mass distribution with a spatial extension and that higher curvature  terms are to be complemented by the Einstein-Hilbert term.
Recently, Schimming and  Schmidt \cite{Schimming:1990jrc}  gave an interesting  historical account of the general quadratic gravity dating back to the seminal work of H. Weyl introducing the concept of the gauge invariance.

The quadratic curvature terms are also popular in cosmological models.  In this context, they offer alternative cosmological models that can  account for the late time accelerated expansion of the universe \cite{Sotiriou:2008rp,carroll2005,NOJIRI20171,NOJIRI201159,PhysRevD.68.123512}.

The field equations for the quadratic curvature gravity were previously studied by Buchdahl \cite{buchdahl1} pointing out that the Palatini procedure with independent connection and the metric theories yield the field equations that are in general different. In contrast,  the Einstein-Hilbert and also the Gauss-Bonnet type higher power Lagrangians yield the second order equations provided that the torsion vanishes by using either of the methods.

The paper is organized as follows. In the following section, we introduce the notation used for the geometrical quantities on  pseudo-Riemannian manifolds. In the ensuing section, we review some identities satisfied by the curvature 2-forms and the irreducible parts of the curvature tensor expressed in the form that is  useful in our presentation. Using the identities introduced, the metric field equations for general quadratic curvature  gravitational Lagrangian are written down. In particular, an analog  of the Bach tensor is introduced by using the curvature identities. In   Sec.~\ref{equivalent-lag}, equivalent forms of the general quadratic curvature lagrangian in four spacetime dimensions are discussed. In Sec.~\ref{action}  the metric field equations are derived by a coframe variation and it is shown explicitly that the variational derivative yields the tensors obtained by the curvature identities. In the ensuing section, the metric field equations for some more general quadratic curvature Lagrangians  are derived.  In Sec.~\ref{lin} the metric field equations are linearized by using perturbation 1-forms. The paper concludes with some remarks on the method and the tensor-valued $p$-forms introduced.

\emph{Notation}. We use the exterior algebra and the tensor-valued forms defined on  pseudo-Riemannian manifolds in our presentation. The exterior $p$-forms are all expressed relative to a set of orthonormal  basis 1-forms $\{e^a\}$ for which the metric can be expressed as
\be
g=\eta_{ab}e^a\ot e^b
\ee
with the constant metric coefficients displayed by the diagonal constant matrix $(\eta_{ab})=diag(-+++)$. The associated frame fields are denoted by $\{X_a\}$ with $g(X_a,X_b)=\eta_{ab}$ and the contraction  operator with respect to the basis frame field is denoted bt $i_a\equiv i_{X_a}$.  Assuming a definite orientation, the invariant volume element is defined as the Hodge dual $*$ of unity as
\be
*1=\frac{1}{4!}\varepsilon_{abcd}e^a\wdg e^b\wdg e^c\wdg e^d
\ee
in terms of the completely antisymmetric permutation symbol $\varepsilon_{abcd}$. The exterior product of basis 1-forms, for example, $e^a\wdg e^b\wdg e^c\wdg e^d$ are often written as $e^{abcd}$ for convenience.

The first structure equations of Cartan  read
\be\label{se1}
de^a+\omega^a_{\fant{a}b}\wdg e^b=0
\ee
in terms of the Levi-Civita connection 1-forms $\omega_{ab}=-\omega_{ab}$ and the exterior derivative of the basis 1-form fields.
The covariant exterior derivative with respect to $\omega_{ab}$ is denoted
by $D$.

The curvature 2-forms $\Omega^{ab}=\frac{1}{2}R^{ab}_{\fant{ab}cd}e^{cd}$ corresponding to the Levi-Civita connection defined in (\ref{se1}) can be obtained by the second structure equations of  Cartan
\be\label{se2}
\Omega^a_{\fant{a}b}=d\omega^a_{\fant{a}b}
+
\omega^a_{\fant{a}c}\wdg \omega^c_{\fant{a}b}.
\ee
The Ricci 1-form $R^a\equiv R^a_{\fant{a}b}e^b= R^{ac}_{\fant{ac}bc}e^b$, and the scalar curvature $R$ are defined in terms of the contractions $R^a=i_b\Omega^{ba}$ and $R=i_aR^a$ respectively. With some nuances, we essentially adopt the notation introduced in Ref. \cite{dereli-tucker2}. The notations for further  geometrical tensors and the tensor-valued forms required in what follows are introduced as the presentation proceeds.

\section{Some Identities Satisfied By the Curvature Tensor}\label{sect-identities}

In relation to an orthonormal basis, Riemann curvature tensor can be broken down into irreducible components \cite{Thirring:1979bh} expressed as
\be\label{curv-decomp}
\Omega^{ab}
=
C^{ab}+D^{ab}+E^{ab}
\ee
where $C^{ab}$ is the Weyl 2-form (corresponding to the conformal curvature tensor), and the second rank part $D^{ab}$ is explicitly given by
\be
D^{ab}
\equiv
\tfrac{1}{2}(e^a\wdg S^b-e^b\wdg S^a)
\ee
in terms of the traceless Ricci 1-forms
$
S^a=R^a-\frac{1}{4}Re^a
$.
 $E^{ab}=\frac{1}{12}Re^{ab}$ represent the trace part expressed in terms of the scalar curvature $R$.
The irreducible components have the same index symmetries as the Riemann tensor $R_{abcd}$.

 In addition to the Hodge dual acting on curvature 2-forms which reads
\be
 *\Omega^{ab}
 =
 \tfrac{1}{2}R^{ab}_{\fant{ab}cd}*e^{cd},
\ee
it is convenient to define the Lie dual \cite{mielke-2017-book,mielke2005GReGr} which is  defined on curvature  2-forms as
 \be\label{right-dual-def}
 \Omega^{ab}*
 \equiv
 \tfrac{1}{2}\epsl^{ab}_{\fant{ab}cd}\Omega^{cd}.
 \ee
The Lie dual operator can also be defined for a general anti-symmetric tensor-valued form. In particular it can be applied  to the irreducible parts of the curvature defined in (\ref{curv-decomp}).
As a consequence of the symmetry of the curvature tensor
$R_{abcd}=R_{cdab}$, the Lie dual defined in (\ref{right-dual-def}) is not independent of the Hodge dual of the corresponding curvature  2-form. In particular, the irreducible parts also have the same index symmetries and consequently, the Hodge duals and the Lie duals of these 2-forms turn out to be the same up to a sign. One can verify that the Hodge duals of the irreducible parts are related to the Lie dual as
 \be\label{double-duality-relations}
 *C^{ab}=C^{ab}*,\qquad *D^{ab}=-D^{ab}*,\qquad *E^{ab}=E^{ab}*.
 \ee
 which are derived previously in \cite{garcia} in the form in (\ref{double-duality-relations}).

 For the Riemann curvature tensor, the second Bianchi identity can be expressed concisely as
\be\label{bi2}
D\Omega^{ab}=0.
\ee
The decomposition in (\ref{curv-decomp}) allows one to rewrite the Bianchi identity (\ref{bi2}) for the curvature form in a convenient form in terms of irreducible parts as
 \be\label{2bi-decomposed}
 DC^{ab}
 =
 \tfrac{1}{2}(e^a\wdg DL^a-e^b\wdg DL^a)
 \ee
with the 1-forms defined in the above equation $L^a\equiv R^a-\frac{1}{6}Re^a$
are known as Schouten 1-form. The covariant exterior derivative of the Schouten 1-forms are known as the Cotton 2-forms denoted by $C^a\equiv DL^a$. The definitions of the Schouten 1-forms and Cotton 2-forms, as well as  the corresponding tensors,  depend explicitly on the number of spacetime dimensions.

The properties of these  tensors were studied extensively by Garcia et al in Ref. \cite{garcia} starting from three spacetime dimensions.
The Cotton and Weyl tensors, and also the related tensors often arise in the discussion of local conformal scaling of the metric by a conformal factor
\be\label{ct}
g\mapsto \bar{g}=e^{2\sigma}g
\ee
where $\sigma$ is assumed to be a function of the local coordinates.
The conformal scaling (\ref{ct}) can conveniently be expressed as the scaling of the basis coframe 1-forms and the contraction with respect to basis frame field as \cite{Dereli:1987ut}
\be\label{coframe-ct}
e^a\mapsto \bar{e}^a=e^{\sigma}e^a,\qquad i_{X_a}\mapsto i_{\bar{X}_a}=e^{-\sigma}i_{X_a},
\ee
respectively. Under the scaling of the metric defined in (\ref{ct}), the Levi-Civita connection transforms inhomogeneously as
\be
\bar{\omega}_{ab}
=
{\omega}_{ab}-X_a(\sigma)e_b+X_b(\sigma)e_a,
\ee
whereas the corresponding curvature 2-forms transform as
\be\label{curvature-ct}
\bar{\Omega}^{ab}
=
{\Omega}^{ab}-e^a\wdg \Sigma^b+e^b\wdg \Sigma^a
\ee
with the 1-form $\Sigma^a$ defined in the above equation having the explicit form
\be
\Sigma^a
\equiv
DX^a(\sigma)-d\sigma X^a(\sigma)
+
\tfrac{1}{2}\eta^{bc}X_b(\sigma)X_c(\sigma)e^a
\ee
in terms of the exterior derivative of the scaling function $X_a(\sigma)\equiv i_ad\sigma$. The formulas (\ref{coframe-ct}) and (\ref{curvature-ct}) allow one to calculate the transformation of a tensor related to the irreducible parts  of curvature tensor under the conformal transformation (\ref{ct}). For example,  one can verify  $\bar{C}^{ab}={C}^{ab}$, and
\be
\bar{R}^a
=
e^{-\sigma}(R^a-2\Sigma^a-\Sigma e^a),\qquad
\bar{R}
=
e^{-2\sigma}(R-6\Sigma)
\ee
where $\Sigma\equiv i_a\Sigma^a$.

An important differential property of the 2-form $C^a$ that was discussed  in \cite{garcia}, which  is useful for the following calculations, is that $i_a C^a=i_aDL^a\equiv 0$ as a consequence of the contracted  Bianchi identity $D\Omega^{a}_{\fant{b}b}\equiv0$, which also implies the relation $i_aDG^a=dR$ in terms of the Einstein form
$G^a=R^a-\frac{1}{2}Re^a$ and the exterior derivative of the scalar curvature $R$.

Yet another curvature identity which turns out to be useful in the following is the relation
\be\label{id0}
 D*E^{ab}
 =
 \tfrac{1}{12}[e^a\wdg *D(Re^b)-e^b\wdg *D(Re^a)]
\ee
which can be verified directly by using the definition $E^{ab}=\frac{1}{12}Re^{ab}$.

Although the exterior covariant derivative does not commute with the Hodge dual $*$ in general, with the help of the double  duality relations in (\ref{double-duality-relations}) together with the second Bianchi identity expressed in the form (\ref{2bi-decomposed}) allow one to write down the identities
 \begin{align}
 D*C^{ab}
 &=
 \tfrac{1}{2}(e^a\wdg *DL^a-e^b\wdg *DL^a),
\label{szekeres-id}\\
 D*D^{ab}
 &=
 \tfrac{1}{2}(e^a\wdg *DR^b-e^b\wdg *DR^a),\label{id2}
 \end{align}
 for the covariant exterior derivatives of the Hodge duals of the irreducible parts $C^{ab}$ and $D^{ab}$, respectively.

The curvature relations (\ref{id0}) and (\ref{id2}) for the irreducible parts $D^{ab}$ and $E^{ab}$ can be combined suitably to obtain, for example,  an equivalent identity as
\be
D*(e^a\wdg R^b-e^b\wdg R^a)
 =
 e^a\wdg *D(R^b+\tfrac{1}{2}Re^b)
 -e^b
 \wdg *D(R^a+\tfrac{1}{2}Re^a).
\ee

The curvature identity in (\ref{szekeres-id}) was introduced in \cite{kundt1,Kundt:2016iky} in a study of exact solutions to the Einstein equations. Later, it was used by Szekeres \cite{szekeres1} in a  different context in an attempt to express the gravitational field equations in a form similar to  the Maxwell's equations, namely   $d*F=-*J$ (in the conventions for the electromagnetic variables adopted in \cite{Thirring:1979bh}) with the matter currents defined by the tensorial expressions on the right-hand side in (\ref{szekeres-id}).

The important differential curvature identity (\ref{szekeres-id}) also occurs in other applications \cite{Choquet-Bruhat:2009xil}, such as the nonlinear stability analysis of the Minkowski spacetime.

In our work, we use an equivalent form of this identity relative to an orthonormal coframe. The Ricci identity of the Hodge dual of the Cotton 2-forms can be written as
\be\label{bach-der1}
D^2 * C^a=\Omega^{a}_{\fant{a}b}\wdg *C^b
=
DL^b \wdg *C^{a}_{\fant{a}b}
\ee
by making use of the symmetry of the inner product provided by the Hodge dual for any two 2-forms. Furthermore, the expression on the right-hand side can be rewritten as
\be\label{bach-der2}
DL_b \wdg *C^{ab}
=
D(L_b \wdg *C^{ab})
+
L_b \wdg D*C^{ab}
\ee
where one can verify that the second term on the right-hand side above vanishes identically with the help of (\ref{szekeres-id}), the first Bianchi identity $R^a\wdg e_a\equiv 0$, and the identity $i_aC^a\equiv 0$. Consequently, the relations (\ref{bach-der1}) and (\ref{bach-der2}) can be used to obtain
\be
D(D*C^a+R_b\wdg *C^{ba})\equiv0.
\ee
The expression in the parenthesis above equation is known as the Hodge dual of the Bach 1-forms $B^a=B^a_{\fant{a}b}e^b$, corresponding to the Bach tensor  $B\equiv B_{ab}e^a\ot e^b$ \cite{bach}.
In our notation, it is convenient to define the Bach tensor by the relation
\be\label{bach-def2}
*B^a
=
D*C^a+R_b\wdg *C^{ba}
\ee
and that it is well-known that the relation $D*B^a\equiv0$ derived above implies that it can also be derived from a variational derivative of the lagrangian $L_W$  (eqn. \ref{qc-equivalence} below) defined in the ensuing section.

To define an analog of the Bach tensor for the trace part in the same manner as above, let us consider the Ricci identity involving the scalar curvature
\be
D^2*D(Re^a)
=
\Omega^{a}_{\fant{a}b}\wdg *D(Re^b)
\ee
where the expression on the right-hand side can be rewritten in the form
\be
D^2*D(Re^a)
=
D[Re_b\wdg *(D^{ab}+E^{ab})]+Re_b\wdg D*(D^{ab}+E^{ab}).
\ee
The higher derivative terms on both sides can be combined to have the expression
\be
D[D*D(Re^a)-Re_b\wdg *(D^{ab}+E^{ab})]
=
Re_b\wdg D*(D^{ab}+E^{ab}),
\ee
and finally, one can show that the term on the right-hand side vanishes identically, whereas the expression on the left-hand side can further be reduced  to have the result as
\be
D[D*D(Re^a)-R*S^a]=0.
\ee
The expression in the above result allows one to define the Hodge dual of  symmetric tensor-valued 1-forms $A_a=A_{ab}e^b$  advertised as the analog of the Bach form in (\ref{bach-def2}) as
\be\label{bach-analogue}
*A^a
\equiv
D*D(Re^a)-R*S^a.
\ee
The curvature squared terms in the definition of the $*A^a$ tensor can also be expressed in the form
\be
R*S^a=-Re_b\wdg *D^{ba}
\ee
involving a particular contraction of the irreducible parts in analogy with the definition of the Bach tensor.
Consequently, the analogy between the two tensors can be exhibited  by rewriting them in the following  forms
\begin{align}
 *A^a&=D*D(Re^a)+Re_b\wdg *D^{ba},
 \\
 *B^a&=D*DL^a+L_b\wdg *C^{ba}.
\end{align}
However, note that $*A^a$ has a trace  given by $e^a\wdg *A_a=-3d*dR$ leading to the well-known massive scalar mode in general quadratic curvature gravity models.

Because the new form satisfies $D*A^a\equiv 0$ by construction, it can be derived from a Lagrangian, namely the quadratic curvature Lagrangian $\frac{1}{4}R^2*1$,   by a variational derivative with respect to the basis coframe  1-forms.

Furthermore, the metric field equations for the most general quadratic curvature gravity can be expressed as a linear combination of the tensors  $*A^a$ and $*B^a$. Before  verifying this claim, it is appropriate to discuss different parameterizations of gravitational Lagrangians involving quadratic curvature terms.

\section{Equivalent Quadratic Curvature Lagrangians}
\label{equivalent-lag}

Four spacetime dimensions turn out to be  special in the context of the quadratic curvature gravity models. This is because of the Euler-Poincar\'{e} four-form which is the quadratic curvature lagrangian defined as
\be\label{EP-def}
L_{EP}
=
\tfrac{1}{2}\Omega_{ab}\wdg \Omega_{cd}*e^{abcd}
\ee
is a total derivative \cite{thirring-wallner,stephenson1958NCim} and consequently, it does not contribute to the gravitational field equations. The particular Euler-Poincar\'{e} form in (\ref{EP-def}) is also known as Gauss-Bonnet form.

By making use of the decomposition (\ref{curv-decomp}), the topological form in (\ref{EP-def}) can be expressed in terms of the irreducible parts as
\be
L_{EP}
=
\tfrac{1}{2}C_{ab}\wdg *C^{ab}
-
\tfrac{1}{2}(R_a\wdg *R^a-\tfrac{1}{3}R^2*1).
\ee
Thus, one can identify the equivalent quadratic curvature lagrangians
\be\label{qc-equivalence}
L_W
\equiv
\tfrac{1}{2}C_{ab}\wdg *C^{ab}
\qquad\text{and}\qquad
L'_W
\equiv
\tfrac{1}{2}(R_a\wdg *R^a-\tfrac{1}{3}R^2*1),
\ee
which differ by the total derivative (\ref{EP-def}).
Moreover, one can also verify, by using the above result that the Riemann tensor-squared lagrangian \cite{yang-1974} can be rewritten in the alternate form
\be\label{riem-squared}
\tfrac{1}{2}
\Omega_{ab}\wdg *\Omega^{ab}
=
R_a\wdg *R^a-\tfrac{1}{4}R^2*1\fant{a}\text{(mod d)}.
\ee

The most general quadratic curvature gravity in four spacetime dimensions can be expressed as a linear combination of the particular Lagrangians $L_W$ and $L_S\equiv \frac{1}{2}R^2*1$, for example, in the form
\be\label{qc-gen1}
L_{QC}
=
\tfrac{\alpha}{2}R^2*1+\tfrac{\beta}{2}R_a\wdg *R^a
\ee
with coupling constants $\alpha, \beta$.
The equivalence of the Lagrangians expressed in (\ref{qc-equivalence}) allows one to rewrite
the lagrangian in (\ref{qc-gen1}) in the form
\be
L_{QC}
=
\tfrac{1}{2}(\alpha+\tfrac{\beta}{3})R^2*1
+
\tfrac{\beta}{2}C_{ab}\wdg *C^{ab}
\ee
with shifted coupling constant for the first term.

The metric field equations that are obtained by a coframe variation of the lagrangian $L_{QC}$ can be expressed in the form
$
*\mathcal{E}^a
\equiv
\delta L_{QC}/\delta e_a
$
in terms of 1-form $\mathcal{E}_a\equiv \mathcal{E}_{ab}e^b.$
The form $*\mathcal{E}^a$ is a linear combination of the Bach form (\ref{bach-def2}) and the new form defined in (\ref{bach-analogue}) as
\be\label{b+a-form}
*\mathcal{E}^a=\beta *B^a+2(\alpha+\tfrac{\beta}{3})*A^a.
\ee
The fourth-order metric field equations for $L_{QC}$ is then given by
$*\mathcal{E}^a=0$ with the explicit expression for $*\mathcal{E}^a$ as
\be\label{gen-qc-eqn}
*\mathcal{E}^a=
D*D[\beta R^a+(2\alpha+\tfrac{\beta}{2})Re^a]
+\beta R_b\wdg *C^{ba}-2(\alpha+\tfrac{\beta}{3})R*S^a.
\ee
In the following section, we verify that the equations (\ref{gen-qc-eqn}) indeed follow from variational derivative of $L_{QC}$.

Before closing the discussion of the equivalent Lagrangians, we note that there is yet another topological lagrangian known as the Chern-Simons term, of the explicit form $\Omega_{ab}\wdg \Omega^{ab}=C_{ab}\wdg C^{ab}$ quadratic in the Riemann curvature tensor \cite{cs,bonilla}. To contribute to the metric field equations, one can couple the term to a dynamic scalar field and we shall discuss this particular lagrangian elsewhere \cite{prep}.
Previously, Dereli et al \cite{Dereli:1987ut}  introduced the locally scale invariant, symmetric tensors in $D=4n-1$ dimensions that are constructed from the coframe variational derivative of Chern-Simons terms.

The curvature tensor satisfying the duality relation of the form $\Omega^{ab}=\Omega^{ab}*$ renders the two topological terms equivalent, $L_{EP}=\Omega^{ab}\wdg\Omega_{ab}$. On the other hand, the curvature tensor satisfying duality relation $\Omega^{ab}*=*\Omega^{ab}$ renders the lagrangian $L_{EP}=\Omega^{ab}\wdg*\Omega_{ab}$ which can also  be expressed in the form given in (\ref{riem-squared}) in terms of irreducible parts. In the latter case, the quadratic curvature models based on the particular form in (\ref{riem-squared}) is bound to yield a zero total energy \cite{deser-tekin1,deser-tekin2} because the lagrangian becomes a total derivative.

\section{Coframe Variation of the Action}\label{action}

The field equations of a model defined by some lagrangian can be obtained by the stationary action principle, which amounts to the extremization of the action integral
\be
I[g]=\int_M L
\ee
with $M$ being a compact manifold possibly with a  boundary $\partial M$ and
$L=L[g]$ stands for a suitable Lagrangian 4-form (or the action density) for the dynamical variables \cite{stephenson1958NCim,thirring-wallner,dereli-tucker2,Kopczynski:1990af,Hehl:1994ue,buchdahl1}. In our case, the metric $g=\eta_{ab} e^a\ot e^b$, or equivalently, the set of the basis coframe forms $\{e^a\}$, are the only dynamical variables  in the gravitational  Lagrangians considered below.

Because connection 1-form given by  $\omega_{ab}$ is not a tensorial geometrical object, it enters into a gravitational Lagrangian 4-form through  corresponding curvature terms in the Lagrangian 4-form in our analysis. Thus, it may be advantageous to consider a given gravitational lagrangian as a functional of the independent set of 1-form variables $\{\omega_{ab}\}$ and $\{e^a\}$ which allow one to calculate variations straightforwardly, and subsequently restrict $\{\omega_{ab}\}$ to be Levi-Civita connection, for example, by introducing Lagrange multipliers to impose the independent connection to be a Levi-Civita connection \cite{dereli-tucker2}.
On the other hand, one can also introduce, for example,  torsional degrees of freedom  and independent field equations for the connection by keeping a general independent connection and the corresponding curvature.

In our notation, the extremization $\delta I=0$ with respect to the coframe yields a 3-form $*\mathcal{E}^a\equiv \mathcal{E}^{a}_{\fant{a}b}*e^b$ obtained by
\be\label{gen-coframe-var-der}
\delta L=\delta e^a\wdg* \mathcal{E}_a=0
\ee
for arbitrary variation $\delta e^a$ of the basis coframe 1-forms. For example, under an infinitesimal local Lorentz transformation,
they take the form \cite{Obukhov:2022xsz}
\be\label{local-frame-rot}
\delta e^a\equiv\epsilon^a_{\fant{a}b}e^b
\ee
with the small parameters $\epsilon_{ab}=\eta_{ac}\epsilon^c_{\fant{a}b} $ satisfying $\epsilon_{ab}+\epsilon_{ba}=0$. Thus, by making use of this expression in the general result (\ref{gen-coframe-var-der}), the invariance of the field equations under infinitesimal rotations in (\ref{local-frame-rot}) implies that
\be\label{local-frame-rot2}
e^a\wdg *\mathcal{E}^b-e^b\wdg *\mathcal{E}^a=0.
\ee
Expressed in a component  form, eqs. (\ref{local-frame-rot2}) imply the symmetry of the indices $\mathcal{E}_{ab}=\mathcal{E}_{ba}$ for any gravitational model with local Lorentz symmetry.

Likewise, for the general covariance of  a Lagrangian 4-form, the variation $\delta e^a$ is to be replaced by the Lie derivative $L_Ve^a$ of the coframe forms with respect to a vector field $V=V^aX_a$. In this case, the general variation (\ref{gen-coframe-var-der}) yields
\be
\delta L=V^a \left[D* \mathcal{E}_a-\omega_{bc}(X_a)(e^b\wdg *\mathcal{E}^c)\right]=0
\ee
which implies the aforementioned diffeomorphism invariance property of a general Lagrangian $D* \mathcal{E}_a=0$ with the help of the local Lorentz invariance and the metricity constraint $\omega_{ab}+\omega_{ba}=0$.

Now we proceed to derive the field equations by a coframe variational derivative. For this,    we adopt a somewhat direct approach \cite{stephenson1958NCim,baykal-ejp-plus}, and subsequently convert the connection variation terms into the coframe variation terms by making use of an equation relating  the  variational expressions for these variables.

The variation of the quadratic curvature lagrangian $L_{QC}$ can be written in the form
\be
\delta L_{QC}
=
\alpha [
2R(\delta R)*1
+
\tfrac{1}{2}R^2\delta*1]
+
\beta[(\delta R_{ab})R^{ab}*1+\tfrac{1}{2}R_{ab}R^{ab}\delta*1].
\ee
To include the curvature 2-forms in the variational procedure, it is convenient to introduce the contractions
\be\label{curv-contractions}
R*1=\Omega_{ab}\wdg *e^{ab},\qquad R_{ab}*1=\Omega_{ac}\wdg *e_{b}^{\fant{a}c}
\ee
in place of the Ricci curvature components $R_{ab}$ and the scalar curvature $R$, and rewrite the total variation in terms of the contractions
in (\ref{curv-contractions}) as
\be
\delta L_{QC}
=
2\alpha R
\delta(\Omega_{ab}\wdg *e^{ab})
+
\beta R^{ab}\delta (\Omega_{ac}\wdg *e_{b}^{\fant{a}c})
-
\delta e^a \wdg i_aL_{QC}.
\ee
With the help of the structure equation (\ref{se2}), the variation of the curvature 2-forms can be expressed in terms of the covariant exterior derivative of the connection 1-forms variation as
\be
\delta\Omega_{ab}=D\delta\omega_{ab}
\ee
where the covariant derivative is with respect to the Levi-Civita connection.
Up to a total derivative, the variational derivative takes the form
\begin{align}
\delta L_{QC}
=&
\delta\omega_{ab}
\wdg
D*\left[
2\alpha Re^{ab}
+
\tfrac{\beta}{2}(e^a\wdg R^b-e^b\wdg R^a)
\right]
\nonumber\\
&+
\delta e^a
\wdg
\left[
\Omega_{bc} \wdg i_a (2\alpha Re^{bc}
+
\beta e^b\wdg R^c)
-
i_aL_{QC}
\right]\qquad{(mod\phantom{a} d)}.
\end{align}
With the help of the curvature identities (\ref{szekeres-id}) and (\ref{id2}) and also taking the metricity constraint $\delta\omega_{(ab)}=0$ into account the variation takes the form
\begin{align}
\delta L_{QC}
=&
-\delta\omega_{ab}\wdg e^b
\wdg
*D\left[\beta R^a+
(2\alpha+\tfrac{\beta}{2})Re^{a}
\right]
\nonumber\\
&+
\delta e^a
\wdg
\left[
\Omega_{bc} \wdg i_a (2\alpha Re^{bc}
+
\beta e^b\wdg R^c)
-
i_aL_{QC}
\right]\qquad{(mod\phantom{a} d)}.
\end{align}
On the other hand, the connection and the coframe variations are not independent since the connection is derived from a metric. Thus, by making use of the total variation of the structure equation (\ref{se1}), one can show that  the $\delta e^a$  and $\delta\omega^a_{\fant{a}b}$ variations are related by the 2-form equations
\be\label{variations-relation}
D\delta e^a+\delta\omega^a_{\fant{a}b}\wdg e^b=0.
\ee
Consequently, with the help of the relations (\ref{variations-relation}), the coframe variations $*\mathcal{E}^a\equiv \delta L_{QC}/\delta e_a$ reduce to the form
\be\label{gen-qc-eqn-form2}
*\mathcal{E}_a
=
D*D\left[\beta R_a+
(2\alpha+\tfrac{\beta}{2})Re_{a}
\right]
+
\Omega_{bc} \wdg i_a (2\alpha Re^{bc}
+
\beta e^b\wdg R^c)
-
i_aL_{QC}.
\ee
Although the coframe variation of $L_{QC}$ does not yield an expression in a form manifestly equivalent to the one in (\ref{gen-qc-eqn}) obtained above, one can insert the irreducible decomposition (\ref{curv-decomp}) into
(\ref{gen-qc-eqn-form2}) and with the help of the first Bianchi identity, one can show that the quadratic curvature terms in (\ref{gen-qc-eqn-form2}) reduce to the form given in (\ref{gen-qc-eqn}).

The restriction of the values of coupling constants by the relation
$\beta+3\alpha=0$,  one arrives at the field equations of the Lagrangian $L_W$ (see, for example, Refs. \cite{garcia,Dereli_1982}), in other words, the form $*\mathcal{E}^a$ for general quadratic curvature gravity in (\ref{gen-qc-eqn-form2}) reduces to the Bach form $*B^a$ in (\ref{bach-def2}).

\section{A General Lagrangian of the Form {$f(R,P)$}}

In applications to cosmology (for example, see the review article \cite{Sotiriou:2008rp}), one often considers   some particular form of the Lagrangian of generic type
\be
L^{gen.}=\tfrac{1}{2}f(R,P)*1
\ee
where the function $f$ is assumed to be a sufficiently differentiable function of its arguments namely the scalar curvature $R$ and the scalar $P$  defined conveniently by the equation
\be
P*1\equiv R_a\wdg *R^a
\ee
as the square of the Ricci tensor.
By proceeding in the same manner as before it is possible to show that the total variation yields a result $*\tilde{E}^a\equiv \delta L^{gen.}/\delta e_a$ that can be expressed in a form similar to those in the previous section as
\begin{align}
 *\tilde{\mathcal{E}}^a=&D*[(df_R+(df_P)_cR^c)\wdg e^a+df_p\wdg R^a]
 +f_PD*[(D(G^a+ Re^a)]
 \nonumber\\
 &+f_PR_b\wdg *C^{ba}-\tfrac{2}{3}f_pR*S^a-2f_R*G^a+\tfrac{1}{2}(f-Rf_R-Pf_P)*e^a=0\label{f-gen-eqn-motion}
\end{align}
where $f_P\equiv \frac{\partial f}{\partial P}$ and $f_R\equiv \frac{\partial f}{\partial R}$.
The last term on the right-hand side is  the Legendre transform of the function $f$ up to a constant. One can verify that the result (\ref{f-gen-eqn-motion}) reduces to the previous tensors, for example to those defined in (\ref{bach-def2}) and (\ref{bach-analogue}).

Although the use of exterior forms allows one to write the metric field equations  in an elegant and easy-to-manipulate form given above, they are considerably complicated even in simple applications. As a sub-case of the above generalized Lagrangian, a simpler Lagrangian 4-form $L=f(P)*1$ is equivalent to a Brans-Dicke type scalar-tensor theory governed by the Lagrangian 4-form as
\be\label{bd-eq.}
L[\phi,g]=\phi R_a\wdg *R^a-V(\phi)*1
\ee
after  the  redefinition of $f_P$ and the potential term for the new scalar field. The equations of motion that follow from (\ref{bd-eq.}) can be obtained from the general form in (\ref{f-gen-eqn-motion}).

More generally, it is possible to find the equations of motion  for a slightly more general $f(P)$-type Lagrangian, for example, of the form
\be\label{gen-f(r,p)-type-lag}
L=f(P,R)*1
\ee
depending on two curvature scalars $R$ and $P$ as defined above, a scalar-tensor equivalent of the Lagrangian in (\ref{gen-f(r,p)-type-lag}) can be written in terms of auxiliary two independent scalar fields $\chi_1$ and $\chi_2$ as
\be\label{gen-f(r,p)-type-eq-lag}
L^{eq.}[\chi_1,\chi_2,g]
=
f(\chi_1,\chi_2)*1+f_{\chi_1}(R-\chi_1)*1+f_{\chi_2}(P-\chi_2)*1
\ee
where the partial derivatives of the function $f$ is written as $\partial f/\partial \chi_i\equiv f_{\chi_i}$ for $i=1,2$. After eliminating the auxiliary fields $\chi_i$ $i=1,2$ by solving the field equations resulting from (\ref{gen-f(r,p)-type-eq-lag}), one  recovers the Lagrangian in (\ref{gen-f(r,p)-type-lag}).

On the other hand, by field redefinitions $f_{\chi_i}\equiv \phi_i$ for $i=1,2,$
the lagrangian (\ref{gen-f(r,p)-type-eq-lag}) can be rewritten in the form
\be\label{gen-f(r,p)-type-eq-lag2}
L^{eq.}[\phi_1,\phi_2,g]
=
\phi_1 R*1+\phi_2 R_a\wdg *R^a-V(\phi_1,\phi_2)*1
\ee
with the potential term given by  the Legendre transform of the original function $f(\chi_1,\chi_2)$ as
\be\label{legendre-tr}
V(\phi_1,\phi_2)
\equiv
\phi_1\chi_1-\phi_2\chi_2-f(\chi_1,\chi_2).
\ee
Consequently, the Lagrangian (\ref{gen-f(r,p)-type-eq-lag2}) is dynamically equivalent to the generic lagrangian in \ref{gen-f(r,p)-type-lag}, provided that   the condition
\be
\left|\frac{\partial \phi_i}{\partial \chi_j}\right|\neq0
\ee
is satisfied. This requirement  ensures that the reduction of the Lagrangian in (\ref{gen-f(r,p)-type-eq-lag}) to the one given in (\ref{gen-f(r,p)-type-lag}), and also ensures that the Legendre transformation defined by (\ref{legendre-tr}) is nonsingular.

Because the scalar fields $\phi_i$ in the Lagrangian (\ref{gen-f(r,p)-type-eq-lag2}) couple to the curvature forms and  therefore couplings of these fields are non-minimal. Because the scalar  fields are dynamical, it is not possible to express the metric field equations  ensuing from (\ref{gen-f(r,p)-type-eq-lag2}) as a linear superposition of the tensors $*A^a$ and $*B^a$ introduced above with constant coefficients.

\section{The Linearized QC Equations}\label{lin}

The use of exterior calculus provides further  insight into the structure  of the quadratic curvature  gravity, for example,  in the context of linearizations of the 3-forms $*A^a$ and $*B^a$  obtained above.

In particular, it is convenient to recall the Sparling-Thirring 2- forms which follow from the attempt to express the Einstein form in a form similar to the Maxwell's  equations expressed in terms of the Faraday 2-form \cite{thirring-wallner}.
To this end, it is convenient to introduce briefly  the linearization of Einstein form $G^a=R^a-\frac{1}{2}Re^a$ defined conveniently by the contraction
of the curvature 2-form as
\be\label{einstein-form-def}
*G^a\equiv-\tfrac{1}{2}\Omega_{bc}\wdg *e^{abc}.
\ee
The expression on the right hand side of (\ref{einstein-form-def})  results from the coframe variational derivative of the Einstein-Hilbert action
\be\label{EH-lag}
L_{EH}=\tfrac{1}{2}R*1=\tfrac{1}{2}\Omega_{ab}\wdg *e^{ab}.
\ee
Explicitly, by introducing the Cartan's structure equations (\ref{se2}) into the right hand side of (\ref{einstein-form-def}), one finds that the Einstein forms split as
\be\label{einstein-split}
*G^a
=
d*F^a-*t^a
\ee
where 2-form $*F^\alpha$, the Sparling-Thirring  forms,   are given by
\be\label{sparling-thirring-forms}
*F^a
\equiv
-\tfrac{1}{2}\omega_{bc}\wdg *e^{abc}
=
e^b\wdg *(de_b\wdg e^a-\tfrac{1}{2}\delta^a_{b}de^c\wdg e_c)
\ee
in terms of coframe basis forms and their derivatives.
The pseudo-tensorial energy momentum forms $*t^a$ can be written, for example, in the form
\be\label{energy-mom-pseudo-tensor-last-form}
*t_a
=
-\tfrac{1}{2}(i_a de^b\wdg F_b-de^b\wdg i_a *F_b).
\ee

For simplicity, let us consider linearization of the metric $g$ around the  flat Minkowski background with the metric of the form
\be\label{flat-metric}
\eta=
\eta_{ab}dx^a\ot dx^b
\ee
in terms of some Cartesian coordinates $\{x^a\}$ with $a=0,1,2,3$. For convenience, the notation  $\{\dot{e}^a\}\equiv \{dx^a\} $ will be employed below.
Instead of introducing  perturbations in metric components, denoted by $h_{ab}$, it is more convenient to use  perturbation 1-forms  defined as  $h_{a}\equiv h_{ab}\dot{e}^b$.
To the first order in the perturbation 1-forms $h_a$,  the general metric tensor takes  the form \cite{Baykal:2016nek}
\be\label{metric-perturbation-def}
g^{(1)}=\eta+\dot{e}^a\ot h_a+h_a\ot \dot{e}^a.
\ee
The perturbation 1-forms  have  symmetric indices $h_{[ab]}=0$ and they are assumed to satisfy  the consistency conditions $|h_{ab}|\ll1$ in the flat background for the consistency of the expansion in (\ref{metric-perturbation-def}).

In component form, the first order metric in terms of perturbation 1-form components, one has $g^{(1)}_{ab}=\eta_{ab}+2h_{ab}$. In the following, the linearization of the geometrical quantities in the metric perturbation is denoted by a superscript $^{(1)}$ or subscript $_{(1)}$ to the quantity.

It readily follows from eqn. (\ref{einstein-split}) that
the linearized Einstein forms can be expressed  in the particular form \cite{thirring-wallner}
\be\label{st-def0}
(*G^a)^{(1)}
=
\star G^a_{(1)}
=
d\star F^a_{(1)}[h]
\ee
where the linearized for of the Sparling-Thirring form (\ref{sparling-thirring-forms})
has the explicit form
\be\label{linearized-ST-form}
\star F^a_{(1)}[h]
=
\dot{e}^b\wdg \star (dh_b\wdg \dot{e}^a)
\ee
and  $\star $ denotes  the Hodge dual operator for the flat background metric
$\eta$. Although it is superfluous, we note that the expression on the right hand side of (\ref{st-def0}) can be rewritten in component form to obtain
the linearized Einstein tensor $\star G^{(1)}_{a}=G^{(1)}_{ab}\star \dot{e}^b$ in the form as
\be\label{lin-eins-tensor}
\star G_{(1)}^{a}
=
-\Box h^a_{\fant{a}b}\star \dot{e}^b
+
\partial_b\partial_c h^{ab}\star\dot{e}^c
+
\partial^a\partial_b h^b_{\fant{a}c}\star\dot{e}^c
-
\partial^b\partial^c h_{bc}\star\dot{e}^a
+
\Box h\star\dot{e}^a
-
\partial^a\partial_b h \star\dot{e}^b
\ee
($\Box\equiv \eta^{ab}\partial_a\partial_b$, $h\equiv h^a_{\fant{a}a}$) by making use of the expression in (\ref{linearized-ST-form}). The equivalent expression for the linearized Einstein form, namely the 3-forms
\be\label{lin-G-form2}
\star G_{(1)}^{a}[h]
=
d[\dot{e}^b\wdg \star (dh_b\wdg \dot{e}^a)]
\ee
facilitates the comparison of the expressions  in terms of the perturbation 1-forms with the corresponding tensor equations in component form.

Moreover, by carrying out the exterior derivatives in the expression (\ref{lin-G-form2}) and evaluating the subsequent inner product, $\dot{e}_b\wdg \star G_{(1)}^{a}[h]=G^a_{(1)b}\star1$, one readily finds the second order partial differential operator corresponding to the linearized Einstein tensor  in a compact and convenient form
\be\label{lin-eins-tensor2}
G^a_{(1)b}[h]
=
\delta^{ace}_{bdf}\partial^d\partial_c h^f_{\fant{a}e}
\ee
in terms of the generalized Kronecker delta symbol $\delta^{ace}_{bdf}$ standing for the determinant
\be
\delta^{ace}_{bdf}
\equiv
\left|
\begin{array}{ccc}
\delta^{a}_{b}&\delta^{a}_{d}&\delta^{a}_{f}\\
\delta^{c}_{b}&\delta^{c}_{d}&\delta^{c}_{f}\\
\delta^{e}_{b}&\delta^{e}_{d}&\delta^{e}_{f}\\
\end{array}
\right|.
\ee
One can readily show that the expression in (\ref{lin-eins-tensor2}) reproduces the expression in (\ref{lin-eins-tensor}), although the equivalent expressions (\ref{st-def0}) and (\ref{lin-G-form2}) are  convenient for practical calculations. Finally,
we note that the linearized Bianchi identity, namely the identity
\be
(D*G^a)^{(1)}=d\star G_{(1)}^{a}[h]\equiv0
\ee
is satisfied as a consequence of the identity $d\circ d\equiv 0$ for the exterior derivative.

The linearized Einstein tensor components $G^{(1)}_{ab}[h]$  displayed in (\ref{lin-eins-tensor}) follow from the Einstein-Hilbert Lagrangian 4-form (\ref{EH-lag}) expressed to second order in the metric perturbation forms $h_a$  that can be written in the explicit form involving a particular kinetic term for the perturbation 1-forms as
\be\label{EH-lag-(2)}
L^{(2)}_{EH}
=
-\tfrac{1}{2}dh_a\wdg \dot{e}^b\wdg \star(dh_b\wdg \dot{e}^a).
\ee
After evaluating the expression on the right hand side by straightforward exterior calculus, the expression  for the  Lagrangian to second order in metric perturbation reduces to
\be\label{EH-lag-(2)-explicit-form}
L^{(2)}_{EH}
=
(
\tfrac{1}{2}
\partial_a h_{bc}\partial^a h^{bc}
-
\partial_a h_{bc}\partial^b h^{ac}
+
\partial_a h^{ab}\partial_b h
-
\tfrac{1}{2}
\partial_a h\partial^a h
)
\star1
\ee
up to an omitted total derivative. Evidently, the components  of the linearized Einstein forms given in (\ref{lin-eins-tensor}) follow from the Lagrangian density given in (\ref{EH-lag-(2)-explicit-form}).
It is worth stressing that, the metric perturbation equations can readily be derived by variating the Lagrangian  (\ref{EH-lag-(2)}) with respect to the perturbation 1-forms as
\be
\delta L^{(2)}_{EH}
=
-\delta h_a\wdg \star G^{a}_{(1)}[h].
\ee

In the same spirit, one can show that
the Pauli-Fierz field equations for a massive spin-2 particle described by a symmetric field $\phi_{ab}$ follow from the Lagrangian 4-form \cite{Fierz:1939ix,hindawi,Hinterbichler:2011tt}
\be\label{PF-lag}
L_{PF}
=
-
\tfrac{1}{2}d\phi_a\wdg \dot{e}^b\wdg \star(d\phi_b\wdg \dot{e}^a)
-
\tfrac{1}{2}m^2\phi_a\wdg \dot{e}^b\wdg \star(\phi_b\wdg \dot{e}^a)
\ee
which  is expressed in terms of the 1-forms fields  $\phi_a\equiv\phi_{ab}\dot{e}^b$, and $m$ denotes the mass of the field.

More recently, the generalized Pauli-Fierz action introduced
in  \cite{deRham:2010ik,deRham:2010kj} involving symmetric polynomial interacting terms for multiple spin-2 fields is studied in \cite{Hinterbichler:2012cn,GrootNibbelink:2006vzk} using the exterior calculus and an extension of the coframe formulation adopted here. For example, the mass term in (\ref{PF-lag}) can be rewritten in the alternate form given in \cite{Hinterbichler:2012cn} by noting that
\be\label{PF-mass-term-form2}
\phi_a\wdg \dot{e}^b\wdg \star(\phi_b\wdg \dot{e}^a)
=
-
\tfrac{1}{2}\varepsilon_{abcd}
\phi^a\wdg \phi^b\wdg \dot{e}^c\wdg \dot{e}^d.
\ee
The particular form of the Pauli-Fierz mass term  in (\ref{PF-mass-term-form2}) allows one to cast the generalized massive spin-2 equations  proposed by de Rham et al \cite{deRham:2010ik} into an equivalent coframe formulation \cite{Hinterbichler:2012cn}.

The mass term in (\ref{PF-lag}) can be motivated by the mathematical consistency for the  Lagrangian. The equations of motion for $\phi_{ab}$ governed by $L_{PF}$ take the form
\be\label{PF-eqns}
\star\mathcal{E}^a_{PF}
=
-
\star G^a[\phi]
-
m^2\dot{e}^b\wdg \star(\phi_b\wdg \dot{e}^a)=0.
\ee
By requiring the divergence of the field equations vanish, $d\star\mathcal{E}^a_{PF}=0$, one  readily ends up with the constraint
\be
d\star(\phi^a-\phi \dot{e}^a)=0,
\ee
whereas the trace of the equations $\mathcal{E}^a_{aPF}=0$ yields
\be
-2 d\star (i_a d\phi^a)-3m^2\phi\star 1=0.
\ee
Thus, one can show that the field equations for $\phi_{ab}$ can be expressed in the form
\be
d\star d\phi^a+m^2\star \phi^a=0
\ee
with the constraints $d\star\phi_a=0$  and $\phi\equiv \phi^a_{\fant{a}a}=0$. These relations imply that  (\ref{PF-eqns}) reduce to the Klein-Gordon type of equations for the components of the perturbation fields $\phi_{ab}$
as
\be
(\Box-m^2)\phi_{ab}=0.
\ee

After this digression on the linearization in terms of perturbation 1-forms suitable for our purposes, let us show that the foregoing  formulas also  apply to linearized QC gravity equations. Remarkably, that the linearized expressions  for both $*A^a$ and $*B^a$ forms can be  expressed in a linearized Sparling-Thirring  2-form by using of the corresponding field equations in the linearized form.

One can make use of the metric field equations given in the form in (\ref{gen-qc-eqn}) to obtain the expression for the linearization of the $*B^a$ form as
\be\label{lin-bach-form-0}
(*B^a)^{(1)}
=
\star B^a_{(1)}
=
d\star F^a[L^{(1)}]
\ee
where the corresponding linearized 2-form belonging to $*B^a$  is given explicitly by
\be\label{lin-bach-form}
\star F^a[L^{(1)}]
\equiv
\dot{e}^b\wdg \star (dL^{(1)}_b\wdg \dot{e}^a)
\ee
in terms of the linearized Ricci curvature 1-forms $L_{(1)}^a=R_{(1)}^a
-
\frac{1}{6}R_{(1)}\dot{e}^a$ displaying a particular nested structure in connection with the original expression given in (\ref{linearized-ST-form}) for the Einstein form. By introducing an  appropriate mass term, the linear superposition of the expressions in (\ref{lin-bach-form}) and in (\ref{linearized-ST-form}) regains the well-known perturbative particle spectrum  including a massive spin-2 particle together with massless spin-2 particle with an infamous relative sign.

The perturbative particle spectrum of the gravity theory defined by the quadratic curvature lagrangian 4-form
\be\label{interacting-massive-spin-2}
L=-L_{EH}+\frac{1}{m^2}L_W
\ee
can also be obtained conveniently by the non-linear metric field equations written in the form $*B^a
+m^2
*G^a
=0$ as
\be\label{massive-spin2-nonlin}
D*DR^a+m^2*R^a=-*R_b\wdg *C^{ba}=0
\ee
where the  quadratic curvature  terms on the right hand side can be considered as  source terms and they are absent in the linear approximation. Together with the identities $D*R^a=R=0$  satisfied, the linearized equations (\ref{massive-spin2-nonlin}) reduce to those of (\ref{PF-lag})
with the identification $\phi_a\equiv R^{(1)}_a$.
Therefore, the metric equations (\ref{massive-spin2-nonlin}) linear in the perturbation fields  then read
\be
(d\star d+m^2)d\star dh_a=[(\Box-m^2)\Box h_{ab}]\star \dot{e}^b=0
\ee
along with the constraints $d\star h^a=0$ and $h=0$.

The connection between the linearized Einstein form $*G^a$ in (\ref{linearized-ST-form}) and the expressions in (\ref{lin-bach-form}) and (\ref{lin-A-form}) for the linearized $*B^a$ and $*A^a$ forms  is also relevant to the discussion of conserved charges that follow from a Lagrangian density of the general form $f(Riemann)$ and the quadratic curvature equivalents of such all-encompassing  higher order theories \cite{Senturk:2012yi,Amsel:2012se} studied previously. The particle content of such theories around their maximally symmetric vacuum solutions is also studied \cite{Tekin:2016vli} determining the masses corresponding to  the spin-0 and spin-2 modes for  the curved background.

The linearized expression for the Bach form (\ref{lin-bach-form-0}) also suggests a method for the construction of the lagrangian 4-form  which is second order in the metric perturbations in terms of the Schouten forms $L^a$ defined above. By inserting the expansion of the curvature form in (\ref{curv-decomp}) into the equation (\ref{einstein-form-def}), one readily ends up with the relation
\be
*G^a
=
L_b\wdg *e^{ba}.
\ee
Moreover, by taking the wedge product of this relation with $L_a$ one finds
the expression
\be
L_a\wdg *G^a
=
L_a\wdg L_b\wdg *e^{ba}
=
L_a\wdg e_b\wdg *(L^{a}\wdg e^{b})
\ee
which has formally the same form as the mass term in the Pauli-Fierz Lagrangian (\ref{PF-lag}).

Furthermore, in terms of the Ricci tensor and the scalar curvature, one can show that the Lagrangian 4-form in (\ref{qc-equivalence}) has yet another equivalent form. For this purpose, let us consider  the Lagrangian with the mass parameter $m$  of the form
\be
L'^{(2)}_W
=
L^{(1)}_a\wdg *G^a_{(1)}[h]
-\tfrac{1}{2}m^2
L^{(1)}_a\wdg e_b\wdg *(L_{(1)}^{b}\wdg e^{a})
\ee
that the linearized Bach forms follow simply replacing curvature 1-forms $L^a$ and $G^a$ by their linearized versions of
$L^a_{(1)}=L^a_{(1)b}\dot{e}^b$ and $G^a_{(1)}=G^a_{(1)b}\dot{e}^b$ respectively.

By making use of the explicit expression (\ref{linearized-ST-form}) for the linearized Sparling-Thirring form, and the fact that the inner product of two forms defined by the Hodge dual is symmetric, one can show that
\be
L_a^{(1)}\wdg \star G^a_{(1)}[h]
=
dL_a^{(1)}\wdg \star F^a_{(1)}[h]
=
h_a\wdg \star G^a_{(1)}[L^{(1)}]\qquad (mod\phantom{a}d)
\ee
with the equality of the Lagrangian terms are assumed to hold up to an  omitted exact form. Consequently, the $h_a$ variables can be eliminated by using the corresponding field equations  which read
\be
\frac{\delta L'^{(2)}_W}{\delta h_a}
=
\star G^a_{(1)}[L^{(1)}]
=
0.
\ee
Consequently, the second order form of the Lagrangian (\ref{interacting-massive-spin-2}) for the  spin-2 component of general quadratic curvature gravity with mass $m$  takes the form
\be\label{massive-spin2-intermediate-lag}
-L^{(2)}_{EH}
+
L'^{(2)}_{W}
=
\tfrac{1}{2}dh_a\wdg \dot{e}^b\star (dh_b\wdg \dot{e}^a)
+
dh_a\wdg \dot{e}^b \star (dL^{(1)}_b\wdg \dot{e}^a)
-
\tfrac{1}{2}
m^2
L^{(1)}_a\wdg e_b\wdg *(L_{(1)}^{b}\wdg e^{a}).
\ee
The addition of $-L^{(2)}_{EH}$ renders the term $h_a\wdg \star G^a_{(1)}[L^{(1)}]$ as a Lagrange multiplier, and consequently, with the help of the
extended Lagrangian, one can identify the auxiliary fields $\pi_{\mu\nu}$ introduced in \cite{hindawi} as the linearized Schouten tensor $L^{(1)}_{\mu\nu}$. The elimination of the perturbation 1-form field $h_a$,  which amounts to the replacement $h_a=-L_a^{(1)}$ in (\ref{massive-spin2-intermediate-lag}) yields the reduced Lagrangian, and one obtains the original Pauli-Fierz Lagrangian (\ref{PF-lag}) for the linearized Schouten 1-form $L^{(1)}_a$.

More generally, the vacuum field equations that follow from (\ref{interacting-massive-spin-2}) can be put in a Maxwell-like form analogous  to the Einstein field equations expressed in the form (\ref{einstein-split}).
Explicitly, one can readily show that the vacuum field equations in this case take the Maxwell-like form as
\be\label{b-split}
*B^a+m^2*G^a
=
d*\mathcal{F}^a+*\mathcal{T}^a=0
\ee
where the generalized superpotential 2-forms $\mathcal{F}^a$ is defined in terms of the Sparling-Thirring form $F^a$ as
\be
*\mathcal{F}^a
\equiv
(*d+m^2)*F^a
\ee
with the additional constraint $i_adF^a=0$.
The corresponding pseudo-energy-momentum forms then take the form
\be\label{def2}
*\mathcal{T}^a
\equiv
m^2
*t^a
-
\left[
d*(\omega^{a}_{\fant{a}b}\wdg R^b)
+
\omega^{a}_{\fant{a}b}\wdg *R^b
+
R_b\wdg *C^{ab}
\right]
\ee
in terms of the mass parameter $m$,  the pseudo-energy-momentum forms $*t^a$ defined in (\ref{einstein-split}) and the Ricci and Weyl curvature tensors.

In the same manner, one can show that for the  spin-0 component of the quadratic curvature gravity equations arising from $R^2*1$ Lagrangian 4-form, the linearized  metric equations take  the form, namely $\star A^a_{(1)}\equiv (*A^a)_{(1)}$. Then the linearized   metric equations can be written as
\be\label{lin-A-form}
\star A^a_{(1)}
=
-\tfrac{1}{2}d\star F_{(1)}^a[R_{(1)}\dot{e}^b]=0.
\ee
 The expression  in equation (\ref{lin-A-form}) follows from the calculation
\be
\star F^a_{(1)}[R_{(1)}\dot{e}^b]
=
\dot{e}^b\wdg \star(d (R^{(1)}\dot{e}_b)\wdg \dot{e}^a)
=
-2\star(d R_{(1)}\wdg \dot{e}^a)
\ee
with the metric perturbation $h_a$  are replaced by the 1-form constructed from
the linearized scalar curvature $R^{(1)}\dot{e}_a$ in the definition given in eqn. (\ref{linearized-ST-form}).

In order to derive the scalar degrees of freedom arising from the $R^2*1$ Lagrangian in parallel to the previous case, it is convenient to  start with the scalar-tensor Lagrangian
\be\label{R-squared-lag-form3}
L_{s}=(R\phi-\tfrac{3}{2}m_0^2\phi^2)*1
\ee
which is equivalent to the pure quadratic Lagrangian
\be
L=\frac{1}{6m_0^2}R^2*1
\ee
the constant $m_0$ is a mass parameter. First,  note that the Lagrangian in (\ref{R-squared-lag-form3}) can be rewritten in the convenient form as
\be
L_{s}=-\phi e_a\wdg *G^a-\tfrac{3}{2}m_0^2\phi^2*1.
\ee
Therefore, the quadratic Lagrangian $L_s$ supplemented with the $L_{EH}$ has the expression
\be\label{R-squared-lag-form4}
L^{(2)}_{EH}
+
L^{(2)}_{s}
=
-
\tfrac{1}{2}dh_a\wdg \dot{e}^b\wdg \star(dh_b\wdg \dot{e}^a)
-
d(\phi \dot{e}_a)\wdg \dot{e}^b\wdg \star(dh_b\wdg \dot{e}^a)
-
\tfrac{3}{2}m_0^2\phi^2\star1
\ee
up to an omitted  total exterior derivative. Consequently, the kinetic term for the scalar field $\phi$ can be  obtained from the metric perturbation 1-forms by the redefinition
\be
h_a\equiv h'_a-\tfrac{1}{2}\phi \dot{e}_a.
\ee
In terms of the perturbation 1-forms $h'_a$, the second order Lagrangian 4-form in (\ref{R-squared-lag-form4}) then takes the form
\be
L^{(2)}_{EH}
+
L^{(2)}_{s}
=
-
\tfrac{1}{2}dh'_a\wdg \dot{e}^b\wdg \star(dh'_b\wdg \dot{e}^a)
-
\tfrac{9}{4}d\phi \wdg \star d\phi
-
\tfrac{3}{2}m_0^2\phi^2\star1
\ee
where the kinetic term for the scalar field decouples from the metric perturbation 1-forms in the new variables.

As in the previous case involving the Bach tensor, the vacuum metric field equations  $*A^a=0$ can also be written in a form analogous to the form in (\ref{einstein-split}) for the Einstein form with the help of the metric field equations in (\ref{bach-analogue}) written out as
\be\label{a-split}
*A^a
=
-
d*[d(*d*i_bF^b)\wdg e^a]
+
\omega^a_{\fant{b}b}\wdg*(dR\wdg e^b)-R*S^a
\ee
involving  the contraction of the Sparling-Thirring forms and the Ricci curvature tensors.

Consequently, the expressions (\ref{b-split}) and (\ref{a-split}) allow one to write generic QC equations in a Maxwell-like form in parallel to the Einstein gravity case. Note that the linearization of the generic quadratic curvature gravity equations can be rewritten in terms of  2-forms (\ref{lin-bach-form}) and (\ref{lin-A-form}) as Sparling-Thirring type 2-forms.

With further development of the above analysis along the lines presented in \cite{Baykal:2018bqd}, the splitting of $*A^a$ and $*B^a$ forms can be performed around a maximally symmetric vacuum solution,  where curved background analogs of the linearized expressions above  can be introduced. Previously, such superpotential forms have been used  in exploring some physical aspects of the critical gravity in four spacetime dimensions \cite{PhysRevLett.106.181302}. By using the time component  of the 2-forms expanded around a maximally symmetric vacuum solution  and the background Killing vector fields, L\"u and Pope calculated the Abbott-Deser-Tekin  mass \cite{deser-tekin2, deser-tekin1} for  anti de Sitter-Schwarzchild black hole solutions to cosmological Einstein gravity complemented with generic QC gravity.

Finally, it is interesting  to compare the Maxwell-like form of the gravitational field equations in this section to the
one proposed by Szekeres \cite{szekeres1} related to the curvature identity (\ref{szekeres-id}).
By defining the right hand side as a two-indexed matter current $*J^{ab}$ with its components  can be determined by the Einstein field equations, the identity (\ref{szekeres-id}) can be rewritten   in the suggestive form
\be\label{szekeres-maxwell-form}
D*C^{ab}
=
*J^{ab}
\ee
with the definition
\be
*J^{ab}
\equiv
\tfrac{1}{2}(e^a\wdg *DL^b-e^b\wdg *DL^a).
\ee
Thus, the decomposition in (\ref{szekeres-maxwell-form}) can be considered as a differential identity relating the free gravitational field represented by  the conformal 2-form $C^{ab}$ with the matter current $J^{ab}$ expressed in terms of  the second rank part of the Riemann curvature tensor analogous to the inhomogeneous Maxwell's equations.

With the help of the Bianchi identities for the curvature 2-form and the curvature decomposition in (\ref{curv-decomp}), one can show that
\be
D^2*C^{ab}
=
\Omega^a_{\fant{a}c}\wdg *C^{cb}
+
\Omega^b_{\fant{a}c}\wdg *C^{ac}
=
0.
\ee
Consequently, the Maxwell-like form of the curvature identity in (\ref{szekeres-maxwell-form}) yields  a covariantly constant matter current satisfying $D*J^{ab}=0$ which explicitly read
\be
e^a\wdg D*DL^b-e^b\wdg D*DL^a=0
\ee
as the gravitational analog of the local charge conservation law $d*J=0$ in terms of the electromagnetic four current 1-form $J$.

The differential identity in (\ref{szekeres-id}) appears in several  equivalent forms in the literature. For example, it can be put in a form as
\be\label{szekeres-id2}
i_aD*C^{ab}=\tfrac{1}{2}*C^b
\ee
by contracting it with $i_a$ and summing over the indice $a$.
In connection with the analogy of the differential curvature identity with  Maxwell's equations, the identity in (\ref{szekeres-id2}) has recently been  introduced by Clark and Tucker \cite{Clark-Tucker2000} in a  study of gravitomagnetism.

\section{Concluding Remarks}\label{conclusion-sect}

The use of the exterior calculus on pseudo-Riemannian manifolds
offers a considerable computational advantage  and flexible treatment of dynamical variables in the variational techniques  involving the first order formalism,  the metric formalism as well as  the variation involving Lagrange multipliers for introducing constrained variables.
Moreover, the coframe variation yields alternate expressions that may provide further insight into the structure of the field equations in general. In this regard, the field equations obtained above can also be expressed relative to a null coframe \cite{Svarc:2022fmg} in  a general survey of algebraically special solutions.

The discussion of the gravitational models confined to four spacetime dimensions can be extended to three spacetime dimensions as well \cite{Dereli:2019ewe,Dereli:2019sgp}. In three spacetime dimensions, there is no fourth-rank irreducible part of the curvature and for this particular reason, the most general form of the quadratic curvature gravity remains as in the four dimensional form given in (\ref{qc-gen1}). However, this is not the case for spacetime dimensions greater than four, for which  the most general form is to include the square of the fourth rank part in addition to $L_{QC}$ in (\ref{qc-gen1}). Thus, it is more appropriate to extend the discussion to higher dimensions by considering  the $C_{ab}\wdg *C^{ab}$, $R_a\wdg*R^a$, and $R^2*1$ terms in place of a term of the form $\Omega^{ab}\wdg *\Omega_{ab}$ because the latter  term can be decomposed into the linear sum of the  terms expressed in the square of the irreducible parts.

\bibliography{QCGR-EQNS-v2}
\bibliographystyle{unsrt}

\end{document}